

\def\yesans{y }
\def\ansr{y }
\ifx\ansr\yesans\message{(The figure will be included.)}
\def\blotz{\input pictex}
\else\message{(The figure will be omitted.)}
\def\blotz{\end}
\fi

\input harvmac
\noblackbox
\def\pr#1#2#3{Phys. Rev. {\bf #1}, #2 (#3)}

\def\plb#1#2#3{Phys. Lett. B {\bf #1}, #2 (#3)}
\def\prd#1#2#3{Phys. Rev. D {\bf #1}, #2 (#3)}
\def\npb#1#2#3{Nucl. Phys. B {\bf #1}, #2 (#3)}
\def\lnc#1#2#3{Lett.\ Nuovo Cimento {\bf #1}, #2 (#3)}
\def\tr{\hbox{tr }}

\Title{\vbox{\baselineskip12pt\hbox{BUHEP-94-10}\hbox{HUTP-94/A014}
\hbox{hep-ph/9406281}}}
{\vbox{\centerline{The Higgs Boson Width is Adjustable}}}

\centerline{R. Sekhar Chivukula$^{a,1}$}
\centerline{Michael J. Dugan$^{a,b,2}$}
\centerline{and}
\centerline{Mitchell Golden$^{b,3}$}
\footnote{}{$^a$Boston University, Department of Physics, 590 Commonwealth
Avenue, Boston, MA 02215}
\footnote{}{$^b$Lyman Laboratory of Physics,
Harvard University, Cambridge, MA 02138}
\footnote{}{$^1$sekhar@abel.bu.edu}
\footnote{}{$^2$dugan@physics.harvard.edu}
\footnote{}{$^3$golden@physics.harvard.edu}

\vskip .4in

\centerline{\bf ABSTRACT}

We show that it is possible to construct models in which the width of
the Higgs boson is arbitrary - either smaller or larger than a
standard model Higgs boson of the same mass.  There are no new fields
into which the Higgs boson decays.  Instead, the coupling of the Higgs
boson to the gauge bosons is adjusted.  We construct and analyze
weakly--coupled models with arbitrary--width Higgs bosons to
investigate the phenomenology one might find in a strongly interacting
model.  In any such model new physics must enter at a mass scale which
decreases as the Higgs boson width is adjusted away from its standard
model value. In particular, if the Higgs boson is wider than the
standard model Higgs boson, then interesting new physics must appear
in the isospin--two channel.

\bigskip

\Date{6/94}

\vfil\eject

\newsec{Introduction}

In the standard model there exists a scalar particle with custodial
isospin zero: the Higgs boson. There is a definite relationship
between the Higgs boson's mass and its width to decay into gauge boson
pairs. Suppose a custodial--isospin--zero scalar is discovered, but with
a width which does not satisfy the standard model relationship.
Though the particle content of the world appears to be the same
as in the standard model, our hint that the standard model is not the
whole story comes only from the unusual size of the coupling of
the Higgs boson to other particles. The scalar particle with isospin
zero in such a model has been termed a non-standard Higgs boson
\ref\rscvk{R.~S.~Chivukula, V.~Koulovassilopoulos,
\plb{309}{371}{1993} \semi V.~Koulovassilopoulos, R.~S.~Chivukula,
Boston University Preprint BUHEP-93-30, Dec 1993, hep-ph - 9312317.}.

In this paper we construct models in which the Higgs boson is narrower
or wider than a standard model Higgs boson of the same mass.  We show
that any theory with a non-standard Higgs particle must contain new
physics at a mass scale which decreases as the couplings of the Higgs
boson deviate from those of the standard model.  We show that if the
Higgs boson is wider than the standard model Higgs boson, and the
model is perturbatively unitary, then the model must contain
isospin--two resonances.

The models we discuss are only meant to be toy examples which contain
a non-standard Higgs boson.  Such a Higgs particle could, for example,
appear in a model with a strongly interacting symmetry breaking
sector.  This strongly interacting symmetry breaking sector, however,
would have to be unlike conventional technicolor \ref\techni{S.
Weinberg, \prd {19}{1979}{1277}\semi L. Susskind, \prd {20} {1979}
{2619}.} which, in analogy to QCD, would be expected to contain only
heavy and broad isospin--zero particles, {\it i.e.}  particles having
masses roughly the same as those of other technicolor resonances and
widths comparable to their masses. The Higgs boson we consider in this
paper is assumed to be light by comparison to any other resonances in
the theory.

The plan of this paper is as follows.  In the next section, we define
notation and construct sample models that include Higgs bosons of
arbitrary widths.  In the third section we show that, if the Higgs
width is larger than in the standard model then perturbative unitarity
implies the existence of isospin--two resonances.  In the fourth
section we use the renormalization group and the triviality of
theories with fundamental scalars to set an upper limit on the mass of
the isospin--two multiplet in a model of fundamental scalars.  This
limit is substantially smaller than that given by the arguments of
section three.  In the fifth section we briefly discuss the
radiative corrections to electroweak parameters in models with Higgs
particles of non-standard widths.  The last section states our
conclusions.

\newsec{How to Vary the Higgs Boson Width}

At sufficiently high energy the scattering of the longitudinally
polarized $W$ and $Z$ bosons is approximately the same as that of the
absorbed Goldstone bosons that would be present in the absence of the
gauging of the electroweak symmetry.  This statement is known as the
``equivalence theorem'' \ref\equivt{J. Cornwall, D. Levin, and
G. Tiktopoulos, \prd{10} {1145} {1974}\semi C. Vayonakis, \lnc{17}
{383} {1976}\semi M.~S.~Chanowitz and M.~K~.Gaillard, \npb{261} {379}
{1985}.}.  Such Goldstone bosons can conveniently be described in the
language of chiral Lagrangians \ref\CCWZ{S.~Coleman, J. Wess, and
B. Zumino, \pr{177}{2239}{1969}\semi C. Callan, S.~Coleman, J. Wess,
and B. Zumino, \pr{177}{2247}{1969}.}.  In our case, the Lagrangian
contains the three ``eaten'' Goldstone bosons and a light isoscalar
Higgs boson \rscvk :
\eqn\chirL{
{\cal L} =
{1 \over 4} \left(v^2 + 2 \xi v H + \xi' H^2 + \ldots\right)
(\tr \partial^\mu \Sigma^\dagger \partial_\mu \Sigma) + \ldots + {\cal L}_H
{}~~,
}
where $\xi$ and $\xi'$ are dimensionless coefficients, $v = 246$ GeV, and
$\Sigma$
contains the swallowed Goldstone bosons $w^a$
\eqn\sigdef{
\Sigma = \exp\left({2i\tau^a w^a \over v}\right)
{}~~.
}
Here $\tau^a$ are the generators of $SU(2)$, normalized to tr $\tau^a \tau^b =
\delta^{ab} / 2$.  The first ellipsis indicates more terms coupling the Higgs
boson to the Goldstone boson, ({\it e.g.} $H^3$, $(\partial H)^2$, etc), and
the second ellipsis indicates terms with more derivatives of the Goldstone
bosons.  These higher order terms are generically suppressed by some large
scale,
denoted $\Lambda_\chi$.  We will assume nothing about $\Lambda_\chi$ except
that it is large compared to the Higgs boson mass.  In \chirL\ ${\cal L}_H$
denotes the self-interaction Lagrangian for the Higgs boson
\eqn\LHdef{
{\cal L}_H = {1 \over 2} (\partial H)^2 - {m_H^2 \over 2} H^2
-{\lambda_3 v \over 3!} H^3 - {\lambda_4 \over 4!}H^4 - \ldots
{}~~.
}
The standard model corresponds to the choice
\eqn\standard{
\xi = \xi' = 1~~,~~\lambda_3 = \lambda_4 = {3m_H^2 \over v^2}
{}~~,
}
with all higher order terms zero.

Using the lagrangian above, we may compute the width of the Higgs
boson into $WW$ and $ZZ$.  At tree-level, for
Higgs bosons with mass much greater than twice the $Z$ mass, the width of the
Higgs boson is
\eqn\Hwid{
\Gamma_H = \xi^2 {3 m_H^3 \over 32 \pi v^2}
{}~~.
}
Including the decay to quarks makes the Higgs boson only slightly
broader, even for a relatively heavy top quark.  In the standard model
where $\xi = 1$, there is a definite relationship between the Higgs
boson's mass and its width.  A measurement of the Higgs boson's width
is a measurement of $\xi$. (The measurement of the parameters $\xi'$,
$\lambda_3$, and $\lambda_4$ is far more difficult than that of $\xi$.
The Higgs boson's width is directly accessible to experiment; its self
coupling is not \rscvk.)

We will now see how the value of $\xi$ can be made different from 1,
by considering models with fundamental scalars.

It is easy to see how one can make $\xi < 1$.  The strength of the
coupling of the Higgs boson to the $W$ and $Z$ is dictated by the need
to give the gauge bosons their observed mass.  To reduce the coupling
of the Higgs boson to the gauge bosons, one divides the vacuum
expectation value among several fields.  The field corresponding to a
mass eigenstate Higgs boson does not get the full vacuum expectation
value, $v$, and so it does not have the usual coupling to the gauge
bosons.  As an example, consider the two--Higgs--doublet model
\foot{For a review of Higgs boson properties in weakly-coupled
theories, see \ref\MSSM{{\it The Higgs Hunter's Guide}, J.~F.~Gunion,
{\it et. al.}, Addison Wesley, New York, 1990.}.}.  The Lagrangian of
the symmetry breaking sector is
\eqn\twodoub{
{\cal L} = (D^\mu \phi_1)^\dagger (D_\mu \phi_1) +
(D^\mu \phi_2)^\dagger (D_\mu \phi_2) - V(\phi_1, \phi_2)
{}~~,
}
where $V(\phi_1,\phi_2)$ is a quartic potential.  The covariant derivative is
\eqn\covder{
D^\mu \phi_i = \left(
\partial^\mu + i g W^\mu \cdot \tau - i{g'\over 2} B^\mu \right) \phi_i
{}~~.
}
If we put
\eqn\separ{
\phi_i =
\left(\matrix{{H_i + v_i + i\phi^0_i \over \sqrt{2}}\cr i \phi^-_i}\right)
}
then the masses of the gauge bosons are, at tree level,
\eqn\MW{
M_W^2 = {g^2 \over 4} (v_1^2 + v_2^2)~~~
M_Z^2 = {g^2 + g'^2\over 4} (v_1^2 + v_2^2)
{}~~,
}
so we conclude that $v_1^2 + v_2^2 = v^2$.  It is customary to define $\beta$
such that $v_1 = v \cos{\beta}$ and $v_2= v\sin{\beta}$. The spectrum of this
model includes two Higgs bosons, a pair of charged pseudoscalars, and one
neutral pseudoscalar.  The mass--eigenstate Higgs bosons are a mixture of
the fields $H_1$ and $H_2$.  We write the lighter Higgs boson as
\eqn\lighthiggs{
H = -\sin\alpha H_1 + \cos \alpha H_2
{}~~.
}
The couplings of these particles to the
gauge bosons may be worked out from \twodoub.  Let us imagine that the lighter
of these two Higgs bosons is observed, but the heavier one eludes
experiment.  The coefficient of the $H W^{a\mu}W^a_\mu$ term in the Lagrangian
is $(1/4) g^2 (-v_1 \sin \alpha + v_2 \cos \alpha)$.  The quantity $\xi$ is the
ratio of this coupling to its value in the standard model, so
\eqn\xires{
\xi = \sin(\beta-\alpha)~~,~~\xi'= 1
{}~~.
}
Therefore, $\xi^2 < 1$.  This result is easily generalized to any
number of doublets.

To get a Higgs boson with $\xi > 1$ requires representations other
than doublets.  On the other hand, we do not want to spoil the
successful tree level prediction of the ratio of the $W$ and $Z$
masses:
\eqn\rhopred{
\rho \equiv {(g^2 + g'^2) M_W^2 \over g^2 M_Z^2} = 1
{}~~.
}
We are led to consider only models that have a ``custodial'' symmetry
\ref\cust{M.~Weinstein, Phys. Rev. {\bf D8}, 2511 (1973)\semi
see also P.~Sikivie, L.~Susskind, M.~Voloshin, and V.~Zakharov,
Nucl. Phys. B.  {\bf 173} (1980) 189.} \techni .  This implies that
the symmetry breaking sector must have a $SU(2)_L \times SU(2)_R
\times U(1)_V$ symmetry\foot{Actually, the lagrangian
\chirL\ written above imposed such a symmetry too.  With only the $SU(2)_W
\times U(1)_Y$ symmetry terms such as tr $\partial^\mu \Sigma^\dagger
\partial_\mu \Sigma T_3$ are not forbidden.}, with the hypercharge embedded
in $SU(2)_R \times U(1)_V$.

Consider an $N \times N$ complex matrix field $\chi_N$ that transforms under
$SU(2)_L \times SU(2)_R$ as
\eqn\chitran{
\chi_N \rightarrow L \chi_N R^{\dagger}~~,
}
where the $L$ and $R$ are $SU(2)_L$ and $SU(2)_R$ group elements in
the spin--$(N-1)/2$ representation. To reduce the
number of fields in the model, we impose the reality condition
\eqn\reality{
V \chi_N V^\dagger = \chi_N^*
{}~~,
}
where $V$ is the $N \times N$ matrix connecting the spin--$(N-1)/2$
representation to its complex conjugate.  (For odd $N$, the representation is
real, so $V$ can be chosen to be the identity.)
We choose hypercharge to be generated by $T_3$ of
$SU(2)_R$, and thus
\eqn\chideriv{
D^\mu \chi_N = \partial^\mu \chi_N + i g W^\mu \cdot T \chi_N - i g'
\chi_N B^\mu T_3 ~~,
}
where $T^a$ are the generators of $SU(2)$ in the spin--$(N-1)/2$
representation. (This is a generalization of the models considered in
\ref\three{ P.~Galison \npb{232}{26}{1984}\semi H. M. Georgi and
M. Machacek, \npb{262}{463}{1985}\semi M. S.  Chanowitz and M. Golden,
\plb{165}{105}{1985}.}.)  The kinetic energy term in the Lagrangian is
\eqn\KE{
{\cal L}_{KE} = {1 \over 2} \tr((D \chi_N)^\dagger (D \chi_N))
{}~~.
}
In order to preserve the relation \rhopred, we assume that the
potential for this field gives it a vev proportional to the identity.
There is one isosinglet field $H$ in $\chi_N$ and we put
\eqn\vevdef{
\chi_N = {H + v_N \over \sqrt{N}} I + \ldots
{}~~.
}
Now the interactions among the gauge and Higgs boson are of the form
\eqn\inter{
{\cal L}_{KE} =
{1 \over 2} {v_N^2 \over N} {\left(1 + {H \over v_N}\right)}^2
(\tr(T_a T_b) g^2 W_a W_b + \tr(T_3^2) (2gg' W_3 B + g'^2 B^2)) + \ldots
{}~~.
}
The $W$ and $Z$ get masses in the correct ratio.  Using $\tr(T_a T_b) =
N(N^2-1)/12$, we find
\eqn\mw{
M_W^2 =  g^2 {N^2 - 1 \over 12} v_N^2
{}~~.
}
 From \inter\ we conclude
\eqn\xiN{
\xi = {v \over v_N} = \sqrt{N^2 - 1 \over 3}
{}~~.
}
Thus we see that the value of $\xi$ can be made as large as we like.

To conclude this section we point out that it is possible to get values of
$\xi$ other than those of \xiN\ by mixing Higgs bosons from different
representations.  If two Higgs bosons with respective couplings to $W$ pairs
parameterized by $\xi_1$ and $\xi_2$ mix by an angle $\alpha$, then the value
of $\xi$ for the lighter of the two Higgs bosons is $\xi = - \xi_1 \sin \alpha
+ \xi_2 \cos \alpha$.

\newsec{Unitarity of Goldstone Boson Scattering}

Let us suppose that an experiment has seen a Higgs boson with width
parameterized by $\xi$.  At what scale need new physics enter, and what
form can it take?

The most general form of the scattering amplitude of the Goldstone bosons
$w^a$ consistent with crossing, Bose symmetry, and custodial isospin
invariance is
\eqn\Adef{
a(w^a w^b \to w^c w^d) =
A(s,t,u) \delta^{ab} \delta^{cd} +
A(t,s,u) \delta^{ac} \delta^{bd} +
A(u,s,t) \delta^{ad} \delta^{bc}
{}~~,
}
where $A$ is a function of the three Mandelstam variables symmetric in its
last two arguments.  The Lagrangian \chirL\ contains two Feynman diagrams that
contribute to $A$ at lowest order in $\Lambda_\chi$: one contact diagram and
one in which the Higgs boson is exchanged in the $s$-channel.  Their sum is
\eqn\Ares{
A(s,t,u) = {s \over v^2} \left(1 - {s \xi^2 \over s-m_H^2}\right)
{}~~.
}
In the standard model in which $\xi = 1$, there is at high energies
a cancellation between the first and second terms.  The growth with
$s$ of the amplitude is truncated by the appearance of the Higgs
boson.  An argument of Lee, Quigg, and Thacker \ref\LQT{B.~W.~Lee,
C.~Quigg, and H.~B.~Thacker, Phys. Rev. {\bf D16}, 1519 (1977).} shows
that if the standard model is to be unitary at tree--level at high
energies, {\it i.e.} if the absolute value of all scattering
amplitudes of all partial waves is to remain less than $1$, then the
mass of the Higgs boson must be less than about 1 TeV.

If $\xi \not = 1$, the absolute value of the amplitude continues to
grow with $s$ even above the Higgs boson mass.  If we assume,
following Lee, Quigg, and Thacker, that the amplitude is to be
unitarized by the exchange of spin--zero resonances, then we see that
there are two possibilities:  resonances of isospin zero
or two.  No other scalars can couple to a pair of Goldstone bosons.

The contribution of another isospin--zero scalar is like that of the
lightest Higgs boson, so it makes a negative contribution to the
amplitude at high energies.  If all the resonances are isospin zero
the requirement that the amplitude stop growing at large $s$
implies \ref\guni{
J.~F.~Gunion, R.~Vega, and J.~Wudka, \prd{42}{1673}{1990}}
\eqn\zerorule{
\sum_i \xi_i^2 = 1
{}~~,
}
where $\xi_i$ is defined for each resonance.  Clearly,
if the model contains a Higgs boson with $\xi>1$, there must be something
other than isosinglets to unitarize the amplitude.

An isospin--two multiplet, by contrast, is exchanged in the $s$, $t$, and $u$
channels.  If the multiplet is represented by a symmetric traceless matrix
$S^{ab}$, it may be included into the chiral Lagrangian by adding
\eqn\LS{
{\cal L}_S = {1 \over 4} (\partial^\mu S^{ab}) (\partial_\mu S^{ab})
+ {C \over v} S^{ab} \partial^\mu \pi_a \partial_\mu \pi_b + \ldots
{}~~,
}
where $C$ is a dimensionless coupling constant.  This field makes the
contribution
\eqn\Scontrib{
{C^2 \over v^2} \left[
{2 \over 3} {s^2 \over s-m_2^2} -
\left({t^2 \over t-m_2^2}  + {u^2 \over u-m_2^2}\right)
\right]
{}~~,
}
to $A(s,t,u)$, where $m_2$ is the mass of the isospin--two
representation.  Note that at high energies, where $s$, $t$, and $u$
are much greater than $m_2^2$, the net contribution of the
isospin--two multiplet to $A$ is
\eqn\hecontrib{
{5 C^2 s \over 3 v^2}
{}~~,
}
a positive quantity.

If, as before, we insist that the amplitude stops growing at high
energies, there will be a sum rule connecting the couplings of the
isosinglet Higgs bosons to those of the isospin--two multiplets\foot{This is
a special case of the sum rules derived in \ref\gunii{J.~F.~Gunion,
H.~E.~Haber, and J.~Wudka, \prd{43}{904}{1991}.}.}
\eqn\sumrule{
\sum_i \xi_i^2 = 1 + {5 \over 3} \sum_j C_j^2
{}~~.
}
Here $i$ runs over the Higgs bosons and $j$ runs over the
isospin--two multiplets.  Therefore, if a model has a Higgs boson with
$\xi > 1$ and is unitarized by the exchange of a spin--zero resonance,
it must contain at least one isospin--two representation.

Consider the example given in the last section, in which the symmetry
breaking sector was an $N \times N$ matrix of fundamental scalars.
After the symmetry breaks, the mass eigenstates fall into degenerate
multiplets of custodial isospin --- one multiplet each of custodial
isospin from $N-1$ down to zero.  The three isospin--one particles are
the absorbed degrees of freedom, but the remaining particles are
physical.  For $N > 2$ there is an isospin-two multiplet and $C$ can be
computed explicitly,
\eqn\CN{
C^2 = {(N^2-4) \over 5}
{}~~.
}
With the parameter $\xi$ is given in \xiN, the sum rule \sumrule\ is
satisfied.

The Lee, Quigg, and Thacker argument gives an upper limit on the
mass--scale of new physics, $M$.  For $m_H^2 \ll s \ll M^2$, the
amplitude is growing in magnitude
\eqn\bigs{
A(s,t,u) \sim (1 - \xi^2) {s \over v^2}
{}~~.
}
Since the isospin--zero, spin--zero amplitude
\eqn\azz{
a_{00}(s) = (1 - \xi^2){s \over 16 \pi v^2}
}
must be bounded in magnitude by one, we conclude that
\eqn\mmax{
M^2 \le {16 \pi v^2 \over | \xi^2 - 1 | } =
{(1750 \hbox{ GeV})^2 \over | \xi^2 - 1| }
{}~~.
}
In the $N \times N$ matrix models, where $\xi > 1$, eqn. \mmax\ bounds
the mass of the isospin--two resonance. In a two-doublet Higgs model,
the eqn. \mmax\ gives a bound on the mass of the heavier Higgs boson.

The argument of this section shows that as the parameter $\xi$ is
changed away from its standard model value, the scale of new physics
is reduced.  Even in a strongly interacting model, which will not in
general be unitarized by perturbative physics (and so the sum rule
\sumrule\ might not be valid, or the isospin--two particles might even
be absent), this qualitative feature of \mmax\ should continue to hold.
There is a maximum scale at which some new physics must enter, and
that scale is reduced as the non-standard Higgs boson looks less
like the standard model Higgs boson.

\newsec{Renormalization Group Analysis}

For the scalar model with $N=3$, one finds $\xi^2 = 8/3$ and the Lee, Quigg,
and Thacker bound \mmax\ on $m_2$ is approximately 1400 GeV.  In this
section we will use the renormalization group and the triviality of
theories with fundamental scalars to set an upper limit on the mass of
the isospin--two multiplet in this model that is substantially smaller.

Suppose one Higgs boson with $\xi^2 = 8/3$ and isospin--two scalars
with $C^2 = 1$ are discovered.  If there is a large hierarchy between
the masses of these particles and the masses of other new physics
(such as compositeness effects in the scalars), the model must be
approximately renormalizable.  This world therefore appears to be
described by the model with $\chi_3$, the $3 \times 3$ matrix \three.
On the other hand, any model with fundamental scalars is trivial, and
thus despite its renormalizability the model does not make sense above
a certain energy scale.  We must set the scale of new physics below
that point.  We may compute the $\beta$--function of this model and
integrate to find the Landau pole.  For the isospin--two bosons to be
approximately fundamental, we must insist that their mass be smaller
than some factor times the mass of the Landau pole.\foot{This is
analogous to the Dashen--Neuberger bound for the standard
model \ref\dashen{R.~Dashen and H.~Neuberger, Phys. Rev.  Lett. {\bf
50}, 1897 (1983).}.  A recent review of lattice simulations and
analytic studies of the Higgs sector of the standard model
\ref\boundref{U.~M.~Heller, M.~Klomfass, H.~Neuberger, and P.~Vranas,
\npb{405}{555}{1993}.} reports that $m_H$ must be  less than about 700 GeV.}

The potential for the $3 \times 3$ model is
\eqn\threepot{
V(\chi) =
g_1(\tr \chi_3^t \chi_3 - v_3^2)^2 +
g_2[3 \tr \chi_3^t \chi_3 \chi_3^t \chi_3 - (\tr \chi_3^t \chi_3)^2]
}
The minimum of this potential is at $v_3 I$ if and only if the dimensionless
constants $g_1$ and $g_2$ are positive.  It is the $N=3$ case of a model with
an $O(N)_L \times O(N)_R$ symmetry, the $\beta$-functions for which are
\ref\pisarski{R.~D.~Pisarski and D.~L.~Stein, Phys. Rev. B. {\bf 23},
3549 (1981) and J. Phys. A. {\bf 14}, 3341 (1981).}
\eqn\betaf{
\eqalign{
2\pi^2 \mu {\partial \over \partial \mu} g_1
=&
(N^2+8)g_1^2 + (2N^2+2N-4)g_1g_2 + 2(N^2 + 2N - 4)g_2^2\cr
2\pi^2 \mu {\partial \over \partial \mu} g_2
=&
12 g_1 g_2 + 2(N^2+2N-6)g_2^2
{}~~.
}
}
This potential gives mass to the Higgs boson and the isospin--two particles.
The masses at tree level are
\eqn\masses{
m_H^2 = 8 g_1 v_3^2~~,~~m_2^2 = 8 g_2 v_3^2~~.
}

We now integrate the $\beta$--functions to find the Landau pole.
We define $t = \log(\mu_P / \mu)$, where $\mu_P$ is the value of $\mu$ at the
Landau pole.  The form of the differential equations is such that both $g_1$
and $g_2$ have a pole at the same value of $\mu$.  The pole in $g_1$ goes like
$1/t$, while the pole in $g_2$ goes like
\eqn\gtwopole{
g_2 \sim \left({1 \over t}\right)^{\scriptstyle 12 \over \scriptstyle (N^2+8)}
}
With this information we define the variables
\eqn\xs{
x_1 = {1 \over g_1}
{}~~,~~
x_2 = \left({1 \over g_2}\right)^{\scriptstyle (N^2+8) \over \scriptstyle 12}
}
and integrate the equations for them.

In \fig\contour{The set of points in $g_1$, $g_2$ space with
$t=1$.} we show the set of points in $g_1$, $g_2$ space with $t=1$.
This is the set of coupling constants $g_1(\mu), g_2(\mu)$, such that
the Landau pole is at a mass of $e\mu$ (where $e$ the base of the
natural logarithm).  The equations \betaf\ are such that if
one performs a transformation
\eqn\xforma{
g_1 \to K g_1
{}~~,~~
g_2 \to K g_2
{}~~,~~
t \to {t \over K}
}
for any positive $K$, then the result is still a solution.  Thus, if we know
where the $t=1$ contour is in $g_1$, $g_2$ space, we know where all the other
contours are.

In \contour\ the value of $g_2$ is never bigger than about .75.
Since the mass of the isospin--two multiplet is
\eqn\prescrip{
m_2^2 = 8 g_2(m_2^2) v_3^2  = 3 g_2(m_2^2) v^2
{}~~,
}
we deduce that the model does not make sense if $m_2$ is more than
about 375 GeV!  This upper limit applies at the small $g_1$ end of the
contour, where the Higgs boson is light.  As the Higgs boson gets
heavier, the limit decreases.

In fundamental scalar theories with $N$ larger than 3 or with scalars
in other representations, these limits only get stronger: the extra
fields give additional positive contributions to the
$\beta$-functions, leading to lower values for the Landau pole for a
given value of $g_1$ and $g_2$. We conclude that in models of
fundamental scalars with $\xi >1$, the isospin--two multiplet
must be fairly light.

\newsec{Radiative Corrections}

We have seen that a wide Higgs boson implies the existence of new
physics at a relatively low scale.  The expectation is that the new
physics makes a contribution to electroweak radiative parameters,
which we parameterize in terms of $S$, $T$, and $U$
\ref\rad{M.~E.~Peskin and T.~Takeuchi, Phys.  Rev. {\bf D46}, 381
(1992).}

Consider first the example of the $3 \times 3$ matrix $\chi_3$ \three.
In this model, $S$ is calculable.  The loops of isospin--two particles
give a new contribution.
\eqn\Scalc{
S =
- {5 \over 36 \pi} \left[ {5 \over 2} -
\log\left({m_H^2 \over m_2^2}\right)\right]
+ {1 \over 12 \pi} \log \left( { m_H^2 \over \hbox{1 TeV} ^2} \right)
}
This is a small negative number whenever $m_H < m_2 < 1$ TeV.

It is easy to see, however, that the corrections to $T$ and $U$ are not
calculable in this model\foot{Corrections to $T$ in this model are discussed
in detail in \ref\guniii{J.~Gunion, R.~Vega, and J.~Wudka, \prd {43} {2322}
{1991}.}.  Our point of view is somewhat different.  In \guniii, it was
assumed that the Higgs and isospin--two particles are fundamental, and the
one--loop computations determined the degree of fine--tuning needed to make
$T$ small enough.  We regard the particles as composite, and instead seek to
limit the scale of their compositeness.}.  Electroweak gauge invariance permits
the addition of custodial symmetry violating terms to the Lagrangian
\eqn\viol{
\eqalign{
{\cal L}_{new} = &
   a_1 \tr[ (D^\mu \chi_3)^t (D^\mu \chi_3) T_3^2]
 + a_2 v_3^2 \tr (\chi_3^t \chi_3 T_3^2)
 + a_3 \tr (\chi_3^t \chi_3 \chi_3^t \chi_3 T_3^2)\cr
&+ a_4 \tr (\chi_3^t \chi_3 T_3 \chi_3^t \chi_3 T_3)
 + a_5 | \tr (\chi_3^t \chi_3 (T_1 + i T_2)^2) |^2
{}~~.
}
}
In fact, the addition of these terms to the Lagrangian is actually
{\it required}.  The gauging of only one of the three generators of
the $SU(2)_R$ breaks the custodial symmetry, and these terms are
required as counterterms for diagrams involving the hypercharge gauge
boson.  The $a_2$, $a_3$, $a_4$, and $a_5$ terms cause the vev of the
(3,3) entry of $\chi_3$ to be different from the (1,1) and (2,2)
entries, splitting the tree--level mass of the $W$ and $Z$.  The $a_1$
term directly shifts the mass of the gauge bosons.  The situation is
precisely analogous to the splitting between the $\pi^\pm$ and $\pi^0$
masses generated by photon loops.  While these coefficients are not
formally computable, one may estimate them by dimensional analysis.
One expects that the renormalized value of $a_2$ is of order
\eqn\diman{
a_2 \sim {g'^2 \over 16 \pi^2} {\Lambda^2 \over v_3^2}
}
where $\Lambda$ is the cutoff of the theory.  The $a_1$, $a_3$, and
$a_4$ terms depend only logarithmically on $\Lambda$.  The $a_2$
term leads to
\eqn\Teqn{
T = {2 a_2 v_3^2 \over \alpha m_2^2}
\sim {1 \over 2 \pi \cos^2 \theta_W} {\Lambda^2 \over m_2^2}
{}~~.
}
Since one expects new physics at the scale of the cutoff, if such a model is
to be phenomenologically acceptable, the mass scale of the new physics is
only about a factor of two heavier than the mass of the isospin--two
resonances.

Another possibility is that the isospin--two scalars do not exist as
approximately fundamental particles, and in that case one must continue to use
the chiral Lagrangian \chirL\ up to $\Lambda_\chi$, where some unspecified new
physics appears.  Here, again, loops containing the $B$ gauge boson require
the addition of counterterms that do not respect the custodial symmetry.  Once
again, $T$ and $U$ are not calculable, but there is a significant difference.
There is no custodial--symmetry--violating dimension--two term in the low
energy theory, and thus the corrections to $T$ are not enhanced by a factor of
$\Lambda^2 / v_3^2$.  Unlike the scalar case, $S$ is not calculable either,
due to the presence of counterterms like \ref\chiS{T.~Appelquist, lectures
presented at the 21st Scottish Universities Summer School in Physics, St.
Andrews, Scotland, Aug 10-30, 1980\semi A.~C.~Longhitano,
\npb{188}{118}{1981}\semi M.~Golden and L.~Randall, \npb{361}{3}{1991}\semi
B.~Holdom and J.~Terning, \plb{247}{88}{1990}\semi A.~Dobado, D.~Espriu, and
M.~Herrero, \plb{253}{161}{1991}.}
\eqn\Sterm{
{\cal L}_S =
\tr\bigg(([D^\mu, D^\nu] \Sigma)^\dagger [D^\mu, D^\nu] \Sigma\bigg)
}
As in any strongly interacting theory, one needs information about the physics
at high energies to say more about the radiative parameters.

\newsec{Conclusions}

We have seen that it is possible to construct models in which the
Higgs boson has any desired width, either larger or smaller than in
the standard model.  The less the Higgs boson looks like its standard
model counterpart, the lower the mass scale at which new physics must
enter.  If the theory is tree--level unitary at high energies and the
Higgs boson width is greater than in the standard model, there must be
isospin--two resonances. Furthermore, if these isospin--two particles
are to be considered approximately fundamental they must be light.
One needs information about the physics at high energies to be
definite about the sizes of electroweak radiative corrections, but
they are expected to be significant.

Work on the production and detection of such a non-standard Higgs
boson is in progress \ref\vk{D.~Kominis and V.~Koulovassilopoulos, in
preparation.}.

\bigskip

\noindent
{\bf Acknowledgments}

We would like to thank Dimitris Kominis, Vassilis Koulovassilopoulos,
Ken Lane, and Elizabeth Simmons for discussions and for comments on
the manuscript.  R.S.C.  acknowledges the support of an Alfred
P. Sloan Foundation Fellowship, an NSF Presidential Young Investigator
Award, and a DOE Outstanding Junior Investigator Award.
M.G. acknowledges the support of an NSF National Young Investigator
Award. This work was supported in part under NSF contracts PHY-9218167
and PHY-9057173 and DOE contract DE-FG02-91ER40676, and by funds from
the Texas National Research Laboratory Commission under grants
RGFY93-278 and RGFY93-278B.

\listrefs
\listfigs

\blotz

$$
\beginpicture
\setcoordinatesystem units <3in,3in>
\setplotarea x from 0.00 to 1.25, y from 0.00 to 1.25
\axis bottom label {$g_1$} ticks numbered from 0.00 to 1.25 by 0.25 /
\axis left label {$g_2$} ticks numbered from 0.00 to 1.25 by 0.25 /
\axis top label {$t = 1$ contour} /
\axis right /
\setlinear
\plot
1.142702 0.019946
1.124304 0.039262
1.105963 0.057961
1.087704 0.076060
1.069551 0.093574
1.051524 0.110520
1.033642 0.126915
1.015921 0.142778
0.998375 0.158127
0.981017 0.172980
0.963856 0.187355
0.946900 0.201270
0.930158 0.214744
0.913633 0.227794
0.897326 0.240438
0.881245 0.252693
0.865386 0.264575
0.849754 0.276102
0.834343 0.287287
0.819153 0.298147
0.804183 0.308697
0.789428 0.318950
0.774884 0.328919
0.760550 0.338619
0.746418 0.348061
0.732482 0.357256
0.718745 0.366219
0.705193 0.374958
0.691821 0.383482
0.678631 0.391808
0.665610 0.399939
0.652752 0.407885
0.640055 0.415657
0.627509 0.423260
0.615112 0.430706
0.602858 0.438002
0.590738 0.445153
0.578749 0.452168
0.566882 0.459052
0.555132 0.465811
0.543496 0.472454
0.531965 0.478983
0.520537 0.485408
0.509202 0.491731
0.497961 0.497961
0.486801 0.504097
0.475723 0.510150
0.464717 0.516120
0.453779 0.522012
0.442902 0.527829
0.432089 0.533586
0.421327 0.539274
0.410612 0.544901
0.399941 0.550471
0.389305 0.555985
0.378705 0.561454
0.368129 0.566870
0.357581 0.572249
0.347048 0.577586
0.336529 0.582885
0.326021 0.588157
0.315510 0.593389
0.305001 0.598599
0.294485 0.603784
0.283955 0.608944
0.273411 0.614091
0.262842 0.619218
0.252247 0.624333
0.241617 0.629435
0.230950 0.634530
0.220239 0.639620
0.209476 0.644702
0.198662 0.649794
0.187785 0.654883
0.176840 0.659977
0.165823 0.665080
0.154725 0.670188
0.143543 0.675318
0.132268 0.680458
0.120893 0.685617
0.109412 0.690798
0.097816 0.695995
0.086100 0.701227
0.074254 0.706480
0.062271 0.711760
0.050142 0.717061
0.037859 0.722393
0.025413 0.727725
0.012795 0.733012
/
\endpicture
$$
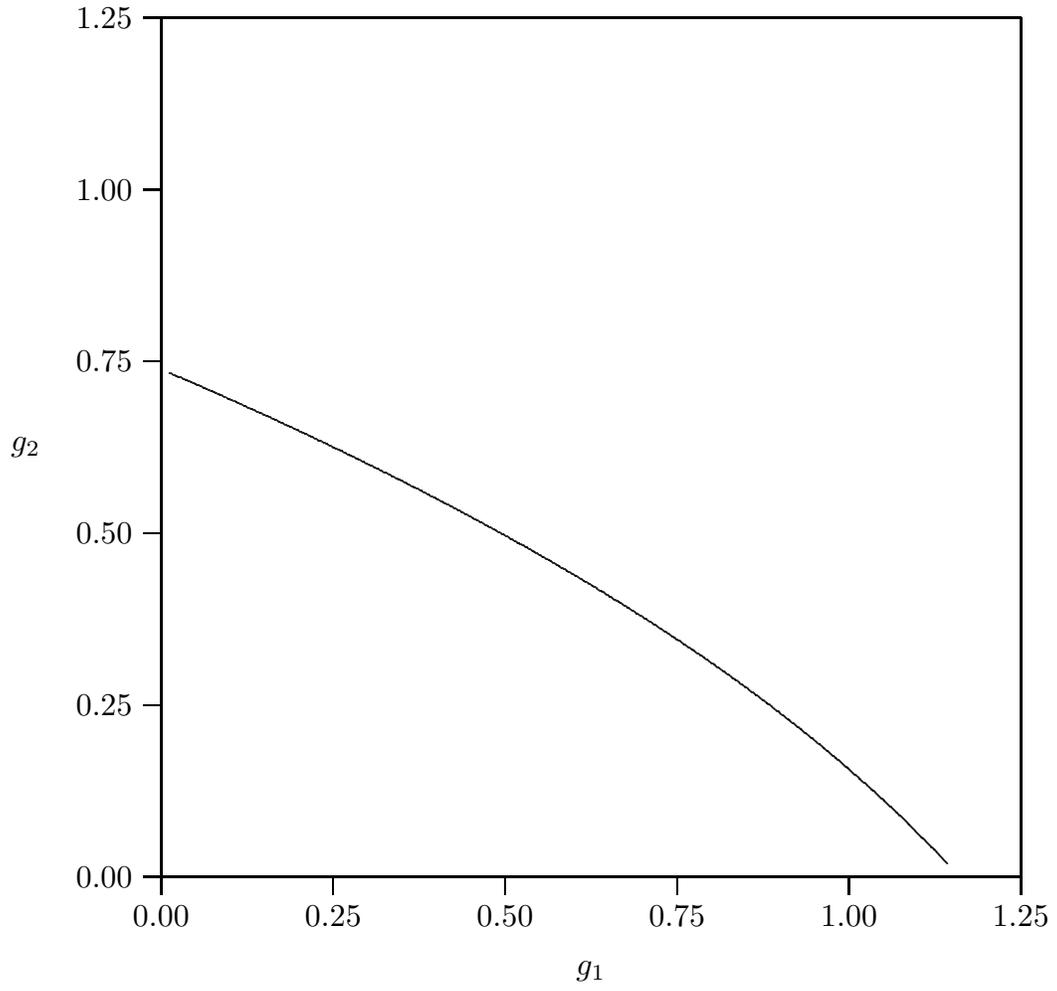
\bye